\documentclass[12pt]{article}
\usepackage{math}
\usepackage{fullpage}
\usepackage{graphicx}
\usepackage{booktabs} 
\usepackage{subcaption} 
\usepackage{epsfig, color}
\usepackage{url}
\usepackage{natbib}
\usepackage[title]{appendix}
\usepackage{setspace}
\graphicspath{ {images/} }
\allowdisplaybreaks

\usepackage{etoolbox}

\newif\ifabbreviation
\pretocmd{\thebibliography}{\abbreviationfalse}{}{}
\AtBeginDocument{\abbreviationtrue}

\begin{document}
\title{Randomization Bias in Field Trials to Evaluate\\ Targeting Methods}
\author{Eric Potash\footnote{Corresponding author. Harris School of Public Policy, The University of Chicago, 1155 E 60th St., Chicago, IL 60637. Email: \url{epotash@uchicago.edu}.}}
\date{}
\maketitle
\begin{abstract}
This paper studies the evaluation of methods for targeting the allocation of limited resources to a high-risk subpopulation.
We consider a randomized controlled trial to measure the difference in efficiency between two targeting methods and show that it is biased.
An alternative, survey-based design is shown to be unbiased.
Both designs are simulated for the evaluation of a policy to target lead hazard investigations using a predictive model. Based on our findings, we advised the Chicago Department of Public Health to use the survey design for their field trial.
Our work anticipates further developments in economics that will be important as predictive modeling becomes an increasingly common policy tool.
\end{abstract}
\clearpage

\section{Introduction}
Policymakers may choose to target the allocation of scarce resources to a subpopulation according to risk or need.
Rapid advances in predictive modeling in recent decades have the potential to make significant contributions to this age-old economic problem \citep{kleinberg2015prediction}.
Some of the programs where predictive targeting is employed or has been proposed include: residential lead hazard investigations \citep{potash2015}, restaurant hygiene inspections \citep{kang2013not}, and violence education \citep{chandler2011predicting}.

Of course, the impact of any targeting method should be evaluated. However, as we shall see, care must be taken in applying the existing economic field trial framework when different treatments (targeting methods) operate on different subsets of the population. We develop a framework for this analysis by drawing on the machine learning \citep{baeza1999modern} and targeted therapies \citep{mandrekar2009clinical} literatures.


Concretely, suppose we have a population of units (e.g. homes) $X=\{1,\cdots,N\}$ and the resources to perform $k$ observations (e.g. investigations) of some binary outcome $y$ (e.g. lead hazards).\footnote{We consider interventions in \S\ref{SectionSurvey}. Continuous outcomes may be accommodated but binary outcomes are more common.}
Next suppose we have a targeting method $S$ which selects a subset $S_k$ of $k$ units for observation.

We define the \textit{precision} of $S$ at $k$ to be the proportion of positive outcomes among the targets $S_k$. When the goal of targeting is to observe positive outcomes, precision is a measure of efficiency (e.g. the proportion of home investigations finding lead hazards). 

In this paper our task is to compare the precision at $k$ of two different targeting methods $S$ and $T$ using $k$ observations. Denoting the precisions of $S$ and $T$ by $\mu_{S_k}$ and $\mu_{T_k}$, respectively, we wish to measure their difference
\begin{equation}
    \delta := \mu_{S_k} - \mu_{T_k}.
\end{equation}
When $\delta$ is positive, $S$ is more efficient than $T$ as a targeting method.

With $k$ observations we can measure the precision of $S$ or of $T$. But we would need up to\footnote{Depending on the size of the intersection $S_k \cap T_k$.} $2k$ observations to measure them both and so measure $\delta$. Thus we estimate $\delta$ statistically.
\begin{figure}
\centering
\begin{subfigure}[t]{0.49\textwidth}
    \centering
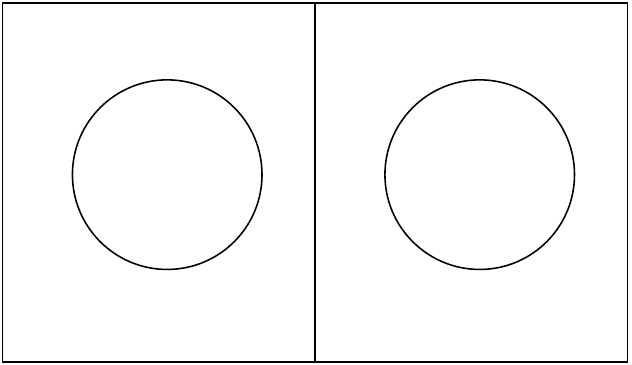
\caption[]{RCT}\label{SubfigureRCT}
\end{subfigure}
\begin{subfigure}[t]{0.49\textwidth}
    \centering
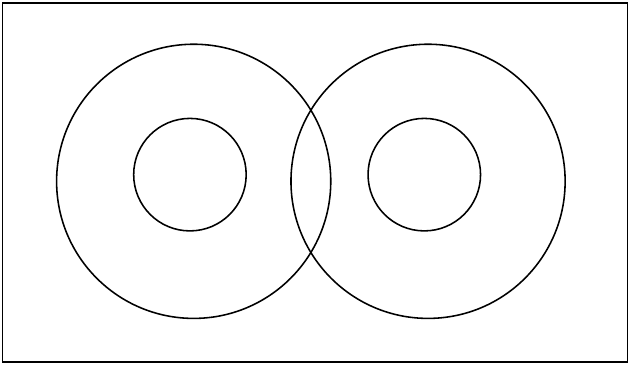

\caption{Survey}\label{SubfigureSurvey}
\end{subfigure}
\caption{In the RCT (\subref{SubfigureRCT}), the population is randomly split in halves $X', X''$ and each targeting method is applied to one half. In the survey (\subref{SubfigureSurvey}), the targeting methods are applied on the population and randomly sampled after excluding their intersection.}\label{FigureDesigns}
\end{figure}

A natural design for a field trial to estimate $\delta$ is an RCT in which the population is randomly split in half and each targeting method is applied to one half. Then we observe the top $k/2$ units in each half, resulting in $k$ total observations.

There is an alternative design: consider $S_k$ and $T_k$ as (after discarding their intersection) disjoint subpopulations and observe $k/2$ random units from each. We think of this design as a survey because it randomly samples the two target sets in the population as opposed to applying the targeting methods to random halves of the population. See figure \ref{FigureDesigns} for a graphical comparison of the two designs.

The remainder of the paper is organized as follows. After defining a framework in \S\ref{SectionFramework}, we show in \S\ref{SectionRCT} that the RCT provides unrepresentative observations and we derive a formula for the bias. 
In \S\ref{SectionSurvey}, we show that the survey gives an unbiased estimate of $\delta$ and discuss implementation details.
In \S\ref{SectionApplication}, we apply the above to a field trial to evaluate targeting of residential lead hazard investigations using a predictive model and simulate sampling distributions for both designs.

The issue in the RCT stems from the interaction between finite populations, partitions, and order statistics.
It is of particular interest as an example of the failure of random assignment to solve an estimation problem.
In this sense it is an example of randomization bias (\citet{heckman1995assessing}, \citet{sianesi2017evidence}) and adds to the collection of pitfalls that researchers should consider before selecting an RCT design \citep{deaton2016understanding}.
Our work anticipates further developments in economics that will be important as predictive modeling becomes an increasingly common policy tool.

\section{Framework}\label{SectionFramework}
A targeting method $S$ is a function which, given a set $X'$ of units and a number $j$ selects a subset $S_j(X')$ of size $j$. When $X'=X$, the full population, we use the shorthand $S_j := S_j(X)$. Let $Y_{S_j(X')}$ denote the outcome $y$ restricted to the set $S_j(X')$ and $\bar Y_{S_j(X')}$ its mean. This is the proportion of units in $S_j$ with positive outcome, i.e. the precision of $S$ at resource level $j$. The population precision at $j$ is then $\bar Y_{S_j}$ but we denote it by $\mu_{S_j}$ to reflect that it is a population (albeit a finite population) object.

Any model of $y$ is also a targeting method. That is, suppose we have such a model which estimates for any unit $x$ the probability\footnote{Or a score which is not necessarily a probability.} $P(y|x)$. The corresponding targeting method would select, from any subset $X'$, the units in $X'$ with the $j$ highest model probabilities.\footnote{Ties may be broken randomly. For simplicity, we do not explicitly consider stochastic targeting methods.}

An expert $S$ may not practically be able to rank all units. Instead, they may only be able to produce a list $S_j(X')$. However, we assume that the expert is rational in the sense that there is an underlying ranking of all units $X$ that is consistently applied to any subset $X'$.This implies that any $S_j(X')$ is ordered and we write
$$S_j(X') = (s_1(X'),s_2(X'),\ldots,s_j(X'))$$
to reference units by their rank. When $X'=X$, we use the shorthand $s_j := s_j(X)$.

\begin{figure}
\centering
	\includegraphics{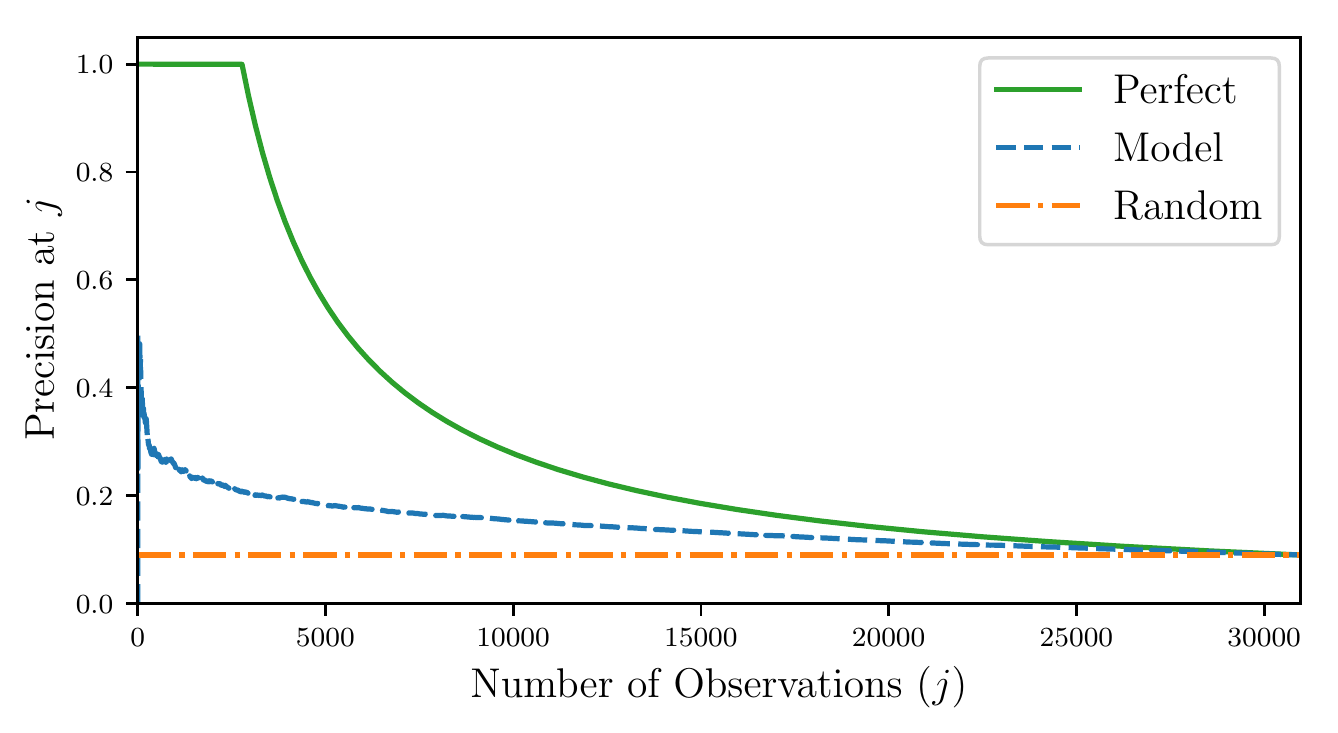}
        \caption{Precision curves for targeting methods in \S\ref{SectionApplication}.}\label{FigurePrecision}
\end{figure}

Following the machine learning literature \citep{baeza1999modern}, we define the \textit{precision curve} of a targeting method $S$ to be $\mu_{S_j}$ as a function of $j$. See figure \ref{FigurePrecision}. Note that when $k=N$ the entire population is selected, so precision at $N$ of any targeting method is the proportion of positive outcomes in the population.

\section{Randomized Controlled Trial Design}\label{SectionRCT}
A natural RCT to estimate $\delta$ using $k$ observations is as follows (see figure \ref{SubfigureRCT}):
\begin{enumerate}
    \item Randomly partition the population into disjoint halves: $X = X' \cup X''$ with $X' \cap X'' = \varnothing$.
    \item Use $S$ to select and observe the top $k/2$ units from $X'$: $S_{k/2}(X')$.
    \item Use $T$ to select and observe the top $k/2$ units from $X''$: $T_{k/2}(X'')$.
    \item Calculate $$\hat\delta_{RCT} := \bar Y_{S_{k/2}(X')} - \bar Y_{T_{k/2}(X'')}.$$
\end{enumerate}
Note we've assumed $N$ and $k$ are even so $N/2$ and $k/2$ are integers.

A hint of the problem with this design arises when carefully defining its terms. Since a traditional RCT applies the same treatment to all units in a treatment group, we must have that: there are just two ``units'', the subpopulations $X'$ and $X''$; the ``treatments'' are $k/2$ selections and observations from each subpopulation; the ``outcome'' is the precision in the subpopulation, e.g. $\bar Y_{S_{k/2}(X')}$. The ``population'' to which $X'$ and $X''$ belong might be the set of all population halves. One quantity, then, that \textit{is} estimated in the RCT is
\begin{equation*}
    \delta_{RCT} := \E\hat\delta_{RCT} = \E[\bar Y_{S_{k/2}(X')} - \bar Y_{T_{k/2}(X')}]
\end{equation*}
where the expectation is taken over halves $X'$.

Besides the fact that the RCT only samples $\delta_{RCT}$ on a single partition, we also argue that $\delta$ rather than $\delta_{RCT}$ is the quantity of interest. This is because $\delta$ measures the difference in the effect of actually implementing either targeting method on the population of interest ($X$) at the scale of interest ($k$). It is not, however, a priori clear what the relationship is between $\delta$ and $\delta_{RCT}$. It might be the case that they are equal.

We now show that this is not the case, i.e. $\delta_{RCT}$ is not in general equal to $\delta$. First consider a single targeting method $S$. Note that the relative top $S_{k/2}(X')$ observed in the RCT is not necessarily a subset of the absolute top $S_k$.
That is, it may contain units ranked beyond the absolute top. 
In fact, when $k \leq N/2$ some halves $X'$ will have relative tops containing \textit{none} of the absolute top.

When $S$ induces a ranking $s_i$, we quantify this by defining $M(X')$ to be the maximum (absolute) rank in the relative top:
\begin{equation}
    M(X') := m \textrm{ such that }s_m = s_{k/2}(X')
\end{equation}
Note that $M \geq k/2$. In appendix \ref{SubsectionRepresentativeness} we show that the distribution on partitions $X'$ induces the following probability distribution on $M$:
\begin{equation}
    P(M=m) = \frac{{{m-1}\choose {k/2-1}}{N-m\choose (N-k)/{2}}}{{N\choose {N}/{2}}}.
\end{equation}
    We marginalize over $M$ in appendix \ref{SubsectionBias} to compute the expected RCT estimate of the precision of $S$ 
\begin{equation}\label{EquationRCTEstimate}
    \E\bar Y_{S_{k/2}(X')} = \sum_{m=1}^N P(M=m)\nu_{S_k, m}
\end{equation}
where $\nu_{S_m,k}$ is a reweighted precision of $S$ at $m$ with increased weight on the last unit
\begin{equation}
    \nu_{S_m, k} := \frac{(k/2-1)\mu_{S_{m-1}} + y_{s_m}}{k/2}.
\end{equation} 

\begin{figure}
    \begin{minipage}[c]{.35\textwidth}
        \includegraphics{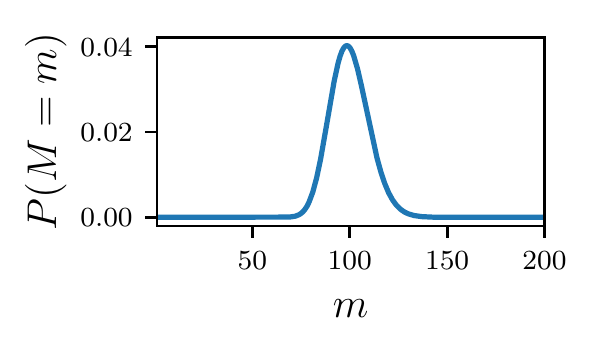}
    \end{minipage}\hfill
    \begin{minipage}[c]{.6\textwidth}
        \caption{The distribution of $M$ when $k=100$ and $N=30,000$ in \S\ref{SectionApplication}. The mode is $k-1=99$.}\label{FigureMpdf}
    \end{minipage}
\end{figure}
We show in \ref{SubsectionRepresentativeness} that $M$ is unimodal with mode at $k-1$.
See figure \ref{FigureMpdf}. Thus the expected RCT estimate of the precision of $S$ is a weighted average over the precision curve.
The greatest weight is placed on $\nu_{S_{k-1}, k}$ with the weight of $\nu_{S_m, k}$ rapidly decaying in the distance from $m$ to $k-1$.

From this we draw several conclusions. First, when the precision is flat (as in the case of random targeting) the RCT estimate is unbiased. Second, (disregarding the difference between $\mu_{S_m}$ and $\nu_{S_m,k}$) bias stems from the difference between the precision curve at $k$ and its value near $k$. However, differences in opposite directions cancel out. Thus bias is especially large near a local extremum. See figure \ref{FigureBiasCurve}. Third, increasing the sample size in the RCT, i.e. going farther down the list in each population half and using $\bar Y_{S_{k/2+1}(X')}$ to estimate $\mu_{S_k}$, does not in general decrease bias.

Finally, combining \ref{EquationRCTEstimate} for both $S$ and $T$, we derive a formula for $\delta_{RCT}$ that is not in general equal to $\delta$. By linearity of expectation we have
\begin{equation}\label{EquationRCTBias}
    \delta_{RCT} = \frac{2}{k}\sum_{m=1}^N P(M=m)(\nu_{S_m,k} - \nu_{T_m,k})
\end{equation}

\section{Survey Design}\label{SectionSurvey}
In the previous section we showed that the RCT observations are not representative of the subpopulations $S_k$ and $T_k$ of interest. We emphasize thinking of $S_k$ and $T_k$ as subpopulations rather than of $S$ and $T$ as treatments. 
Their intersection $I := S_k \cap T_k$ is not necessarily empty. But the units in the intersection are irrelevant to the difference $\delta$:
\begin{equation}\label{EquationMinusIntersection}
    \delta = \alpha(\bar Y_{S_k\backslash I} - \bar Y_{T_k\backslash I}).
\end{equation}
where $\alpha := 1 - |I|/k$.
We use this to design a survey (see figure \subref{SubfigureSurvey}):
\begin{enumerate}
\item Use $S$ to select the top $k$ units from the population: $S_k$.
\item Use $T$ to select the top $k$ units from the population: $T_k$.
\item Observe outcomes for a random sample $S_k'$ of size $k/2$ from $S_k\backslash I$.
\item Observe outcomes for a random sample $T_k'$ of size $k/2$ from $T_k\backslash I$.
\item Estimate $$\hat \delta_{Survey} := \alpha(\bar Y_{S_k'} - \bar Y_{T_k'}).$$ 
\end{enumerate}

The above discussion and the fact that $S_k'$ and $T_k'$ are random samples of $S_k\backslash I$ and $T_k\backslash I$, respectively, implies that $\hat\delta_{Survey}$ is an unbiased estimator of $\delta$:
    $$\E\hat\delta_{Survey} = \delta.$$

Note that if the goal of the trial is only to estimate $\delta$, statistical power is maximized by allocating no observations to the intersection as specified above. On the other hand, to estimate absolute quantities $\mu_{S_k}$, $\mu_{T_k}$ the intersection should be sampled as well. Efficiency could further be increased by stratifying the survey (e.g. across neighborhoods).

Above we have focused exclusively on observation outcomes. When the targets receive an intervention, we may simply use an RCT\footnote{With the usual Stable Unit Treatment Value Assumption (SUTVA)\citep{rubin1980randomization}} within each of the subpopulations $S_k\backslash I$ and $T_k\backslash I$. Then we would estimate the difference in treatment effects $\tau_{S_k} - \tau_{T_k}$ where $\tau := y_1 - y_0$ in a potential outcomes framework.

Policymakers may be uncertain about the resources $k$ for the targeted policy. It is possible for $S$ to outperform $T$ in the top $k_1$ but not in the top $k_2$. Thus, it may be useful select $k$ as an upper bound and sample $S_k$ and $T_k$ with some stratification $k_1 < k_2 < \cdots < k$ to compare the precision of $S$ and $T$ on each risk stratum.

Of course, the usual concerns about generalizability (to other time periods, other populations, etc.) apply as well to this field trial design.


\section{Application}\label{SectionApplication}
\begin{figure}
\centering
\includegraphics{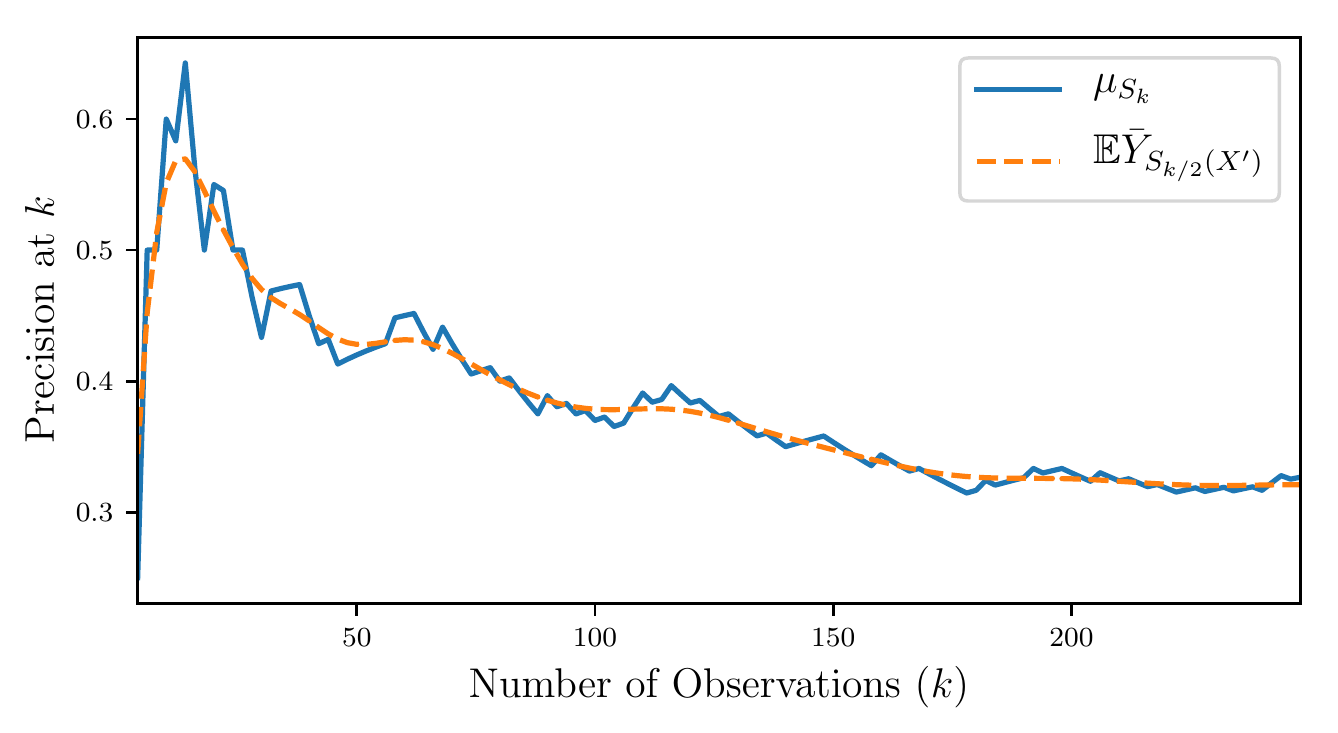}

\caption{True precision $\mu_{S_k}$ and expected RCT estimate $\E \bar Y_{S_{k/2}(X')}$ for $k$ up to 250 in \S\ref{SectionApplication}.
}\label{FigureBiasCurve}
\end{figure}
In \citet{potash2015} we developed a machine learning model to predict which children are at highest risk of lead poisoning using historical blood lead levels, building characteristics, and other data. In this section, we compare the RCT and survey for a field trial to estimate the improvement $\delta$ in precision (i.e. proportion of investigations finding hazards) of targeting $k$ investigations using the predictive model $S$ over random selection $T$.

In this application $N\approx 30,000$ for the population $X$ of a Chicago birth cohort, restricting our attention to children residing in homes built before 1978, the year in which lead-based residential paint was banned \citep{uscprsc1977}. To simulate field trial results we need rankings of children and outcomes of investigations. We take these from \citet[\S7]{potash2015}, which evaluated out-of-sample predictions on the 2011 birth cohort.
Since proactive investigations were not performed, lead hazard outcomes were not available for the population. Instead, blood lead level outcomes were used as a proxy. The resulting precision curve is reproduced in figure \ref{FigurePrecision}.

Using equation \ref{EquationRCTEstimate} we calculated $\E \bar Y_{S_{k/2}(X')}$, the expected RCT estimate of the precision of the predictive model, for $k$ up to 250. These are plotted together with the true precision $\mu_{S_k}$ in figure \ref{FigureBiasCurve}. Their difference is the bias of the estimate and is a function of the shape of the precision curve near $k$. This bias, as a percentage of the true value, varies in this range between -11\% and 9\% with an average magnitude of 2\%.
\begin{figure}
\centering
\begin{subfigure}[c]{0.55\textwidth}
    \centering
\includegraphics{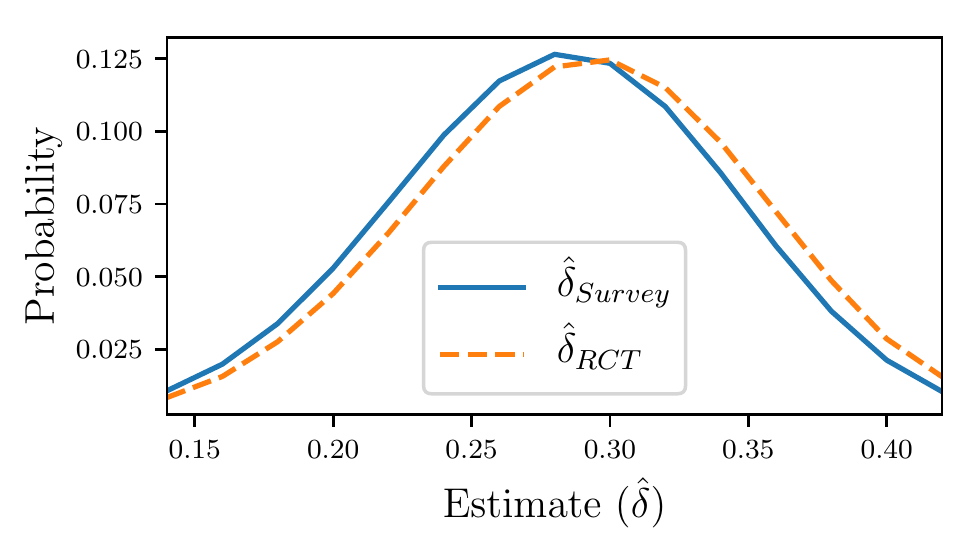}
\end{subfigure}\hfill
\begin{subfigure}[c]{0.42\textwidth}
    \centering
    \begin{tabular}[b]{rll}
    \toprule

    & $\hat\delta_{RCT}$ & $\hat\delta_{Survey}$ \\ \hline
    mean & 0.291 & 0.282 \\ 
    std. dev.& 0.064 & 0.063 \\ 
    bias &0.009 & 0. \\ \bottomrule
  \end{tabular}
\end{subfigure}
\caption{Sampling distribution and summary for estimates from the two designs in \S\ref{SectionApplication}.}\label{FigSimulation}
\end{figure}

Next we estimated full sampling distributions for the RCT and survey designs at $k=100$.
For the RCT, we used Monte Carlo simulation: we calculated $\hat\delta_{RCT}$ over $10^6$ random partitions.
The mean of this empirical distribution agreed to five significant figures with the value of $\delta_{RCT}$ that we derived in equation \ref{EquationRCTBias}.
To estimate the survey results, we computed the distribution of $\hat\delta_{Survey}$ exactly using hypergeometric formulas.

The resulting distributions are displayed and summarized in figure \ref{FigSimulation}.
We find that the RCT is biased to overestimate $\delta$ by 3\%.
It also has higher variance than the survey, which is unbiased as expected.
As discussed in \S\ref{SectionRCT} and illustrated in figure \ref{FigureBiasCurve}, the direction and magnitude of the RCT bias stems from the shape of the precision curve near $k=100$.
In light of these results, we advised the Chicago Department of Public Health to use the survey design for a field trial of a targeted lead investigations policy.
\section{Acknowledgements}
We thank Dan Black, Chris Blattman, Matt Gee, Jesse Naidoo, and the anonymous reviewer for feedback on the manuscript. Thanks for useful discussions to members of the Center for Data Science and Public Policy (DSaPP) and the Energy Policy Institute at Chicago (EPIC).

\pagebreak
\bibliographystyle{apalike}
\bibliography{trial}

\pagebreak
\begin{appendices}
\section{}
\subsection{Relative Top}\label{SubsectionRepresentativeness}
Recall that $M(X')$ is the maximum absolute rank in $S_{k/2}(X')$.
We can write the event $M=m$ as the intersection of two simpler events: of the absolute top $m-1$ exactly $k/2-1$ are in $X'$, i.e. $\#(S_{m-1} \cap X') = k/2-1$; and $s_m$ is in $X'$. Denoting these events by $A$ and $B$, respectively, we derive the distribution of $M$ induced by the distribution on partitions:
\begin{align}
    P(M=m) =& P(A)\cdot P(B|A) \nonumber\\
    =&\textrm{HG}(k/2-1;N,m-1,N/2)\cdot\frac{(N-k)/2+1}{N-m+1} \nonumber \\
    =&\frac{{{m-1}\choose {k/2-1}}{N-m\choose (N-k)/{2}}}{{N\choose {N}/{2}}}
\end{align}
where $\textrm{HG}$ is the hypergeometric distribution.

To find the mode of this distribution we calculate the change between consecutive probabilities:
\begin{align}
    \frac{P(M=m+1)}{P(M=m)} = \frac{m(N/2+k/2-m-1)}{(m-k/2+1)(N-m-1)}
\end{align}
It follows that 
\begin{align}\label{EquationMRatio}
    P(M=m+1) &> P(M=m)\iff m < (k-2)(N-1)/(N-2)\nonumber\\
    P(M=m+1) &= P(M=m)\iff m = (k-2)(N-1)/(N-2)\\
    P(M=m+1) &< P(M=m)\iff m > (k-2)(N-1)/(N-2)\nonumber
\end{align}
Since $m$ is an integer, $N$ is even, and $k \leq N$, equality in \ref{EquationMRatio} implies $k=N$. We conclude that $M$ is unimodal with mode at $k-1$ and probability increasing until that point. If $k < N$ then the probability is decreasing after it.

\subsection{Bias}\label{SubsectionBias}
We can write the relative precision at $k/2$ as a weighted average of the precision of all but the last unit with the outcome of the last unit, which by definition has rank $M$ in the population:
\begin{equation}
\bar Y_{S_{k/2}(X')} = (\bar Y_{S_{k/2-1}(X')}\cdot(k/2-1) + y_{s_M})/(k/2).
\end{equation}
Then the expectation (over partitions) of the relative precision at $k/2$ can be computed conditional on $M$:
\begin{align}
    \E[\bar Y_{S_{k/2}(X')} | M=m] &= (\E[\bar Y_{S_{k/2-1}(X')}| M = m]\cdot (k/2-1) + y_{s_m})/(k/2) \nonumber\\
    &= \left(\E[\textrm{HG}(m-1, (m-1)\mu_{S_{m-1}},k/2-1)] + y_{s_m}\right)/(k/2) \nonumber \\
    &= \left(\left(k/2-1\right)\mu_{S_{m-1}} + y_{s_m} \right)/(k/2).
\end{align}
We define $\nu_{S_m, k}$ to be this quantity, which is the precision at $m$ reweighted.
Marginalizing over $M$, the unconditional expected precision of $S$ in the RCT is
\begin{align}
    \E\bar Y_{S_{k/2}(X')} &= \sum_{m=1}^N P(M=m)\nu_{S_m, k}.
\end{align}

\end{appendices}
\end{document}